# Establishing Direct Phenomenological Connections between Fluid and Structure by the Koopman-Linearly-Time-Invariant Analysis

Cruz Y. Li[12] (李雨桐), Zengshun Chen[1*] (陈增顺), Tim K.T. Tse[3**] (谢锦添), Asiri Umenga Weerasuriya[4], Xuelin Zhang[5] (张雪琳), Yunfei Fu[6] (付云飞), Xisheng Lin[7] (蔺习升)

[1] *Department of Civil Engineering, Chongqing University, Chongqing, China*

[2,3,4,6,7] *Department of Civil and Environmental Engineering, The Hong Kong University of Science and Technology, Hong Kong SAR, China*

[5] *School of Atmospheric Sciences, Sun Yat-sen University, Zhuhai, China.*

[1] yliht@connect.ust.hk; ORCID 0000-0002-9527-4674
[2] zchenba@connect.ust.hk; ORCID 0000-0001-5916-1165
[3] timkttse@ust.hk; ORCID 0000-0002-9678-1037
[4] asiriuw@connect.ust.hk; ORCID 0000-0001-8543-5449
[5] zhangxlin25@mail.sysu.edu.cn; ORCID 0000-0003-3941-4596
[6] yfuar@connect.ust.hk; ORCID 0000-0003-4225-081X
[7] xlinbl@connect.ust.hk; ORCID 0000-0002-1644-8796

[*] Co-first author with equal contribution.

[**] Corresponding author

All correspondence is directed to Dr. Tim K.T. Tse.

In this work, we introduce a novel data-driven formulation, the Koopman-Linearly-Time-Invariant (Koopman-LTI) analysis, for analyzing Fluid-Structure Interactions (FSI). An implementation of the Koopman-LTI on a subcritical free-shear flow over a prism at *Re=22,000* corroborated a configuration-wise universal Koopman system, which approximated the configuration's nonlinear dynamics with stellar accuracy. The Koopman-LTI also successfully decomposed the entwined morphologies of raw measurement into a linear superposition of frequency-based constituents. Most importantly, with random and anisotropic turbulence, the Koopman-LTI yielded frequency-wise identical modes for structure response and fluid excitation, thus establishing direct constitutive relations between the phenomenology of fluid and structure.

Despite centuries of scientific exploration, Fluid-Structure Interaction (FSI) remains an enigma of knowledge. The difficulty of analyzing FSI is trifold. First, a fluid system contains infinite fluid particles that each pertain to a unique motion governed by nonlinear dynamics [1], [2]. Second, turbulence creates highly volatile morphologies, so measured data are in reality an amalgam of thousands of phenomena acting together [3]. Third, the interactive mechanisms of FSI are sophisticated, nonlinear, and deeply esoteric [4]. Without a solution for the Navier-Stokes (NS) equations, the analytical path to tackling FSI is drenched with thorns and obstacles. Nevertheless, recent advancements in computer science luminated the data-driven route [5], [6]. To this end, the Koopman analysis emerges as a brilliant method to overcome the aforementioned difficulties [7]–[10].

By imposing an infinite-dimensional Koopman operator $\boldsymbol{A}$ that maps two data sequences, $\boldsymbol{U_2} = \boldsymbol{AU_1}$, $\boldsymbol{A}$ globally linearizes the nonlinear system. Practically, $\boldsymbol{A}$ is impossible to obtain. However, as proposed by [11], a similarity matrix $\tilde{\boldsymbol{A}}$ can tractably approximate $\boldsymbol{A}$ on a finite-dimensional subspace through an algorithm called the Dynamic Mode Decomposition (DMD) [12]. If $\boldsymbol{U_1}$ and $\boldsymbol{U_2}$ represent a fluid system, $\tilde{\boldsymbol{A}}$ is quintessentially a data-driven, reduced-order, and implicit representation of the Navier-Stokes. The beauty of the Koopman analysis is that prior knowledge is not required to obtain $\tilde{\boldsymbol{A}}$, hence elegantly bypassing the analytical impossibility of the NS. More importantly, a theoretical possibility exists for $\tilde{\boldsymbol{A}}$ to reduce into a linearly time-invariant (LTI) system. With it, applying the eigendecomposition to $\tilde{\boldsymbol{A}}$ can effectively dissect the input data into a configuration-wise universal set of linear constituents:

$$\boldsymbol{U}_{LTI}(\boldsymbol{x},t) = \sum_{i=1}^{n} \alpha_i \phi_i e^{\omega_i t} = \boldsymbol{\alpha \Phi} e^{\boldsymbol{\Omega} t} \tag{1}$$

where $\boldsymbol{U}_{LTI}(\boldsymbol{x},t)$ denotes the reconstructed data at location $\boldsymbol{x}$ and time $t$, $\boldsymbol{\alpha}$ contains the leading coefficients, $\boldsymbol{\Phi}$ contains the eigenfunctions (*i.e.,* DMD modes or Ritz descriptors), and $\boldsymbol{\Omega}$ contains the eigenvalues (*i.e.,* modal frequencies). **Fig.** 1 illustrates the conceptual process of the Koopman-LTI analysis. Its novelty is also hereby clarified: the Koopman-LTI is not a newly invented decomposition technique---it stands on the shoulders of the mathematical giants who have made the Koopman operator theory and the DMD possible [12]–[16]. Instead, it appeals to a brand-new, deliberate rendering of the Koopman analysis. The DMD is employed only as an algorithm to approximate the Koopman modes, which can be substituted by the Koopman Mode Decomposition (KMD) or the Spectral Proper Orthogonal Decomposition (SPOD). Readers are referred to [17] for detailed mathematical formations of the DMD.

This visionary prelude will demonstrate and substantiate the possibility of the Koopman-LTI analysis by successfully 1) constructing an accurate representation of a turbulence-rampant fluid-structure system, 2) decomposing the nonlinearly entwined raw data into distinct linear constituents, and 3) establishing direct phenomenological relationships between the fluid and structure.

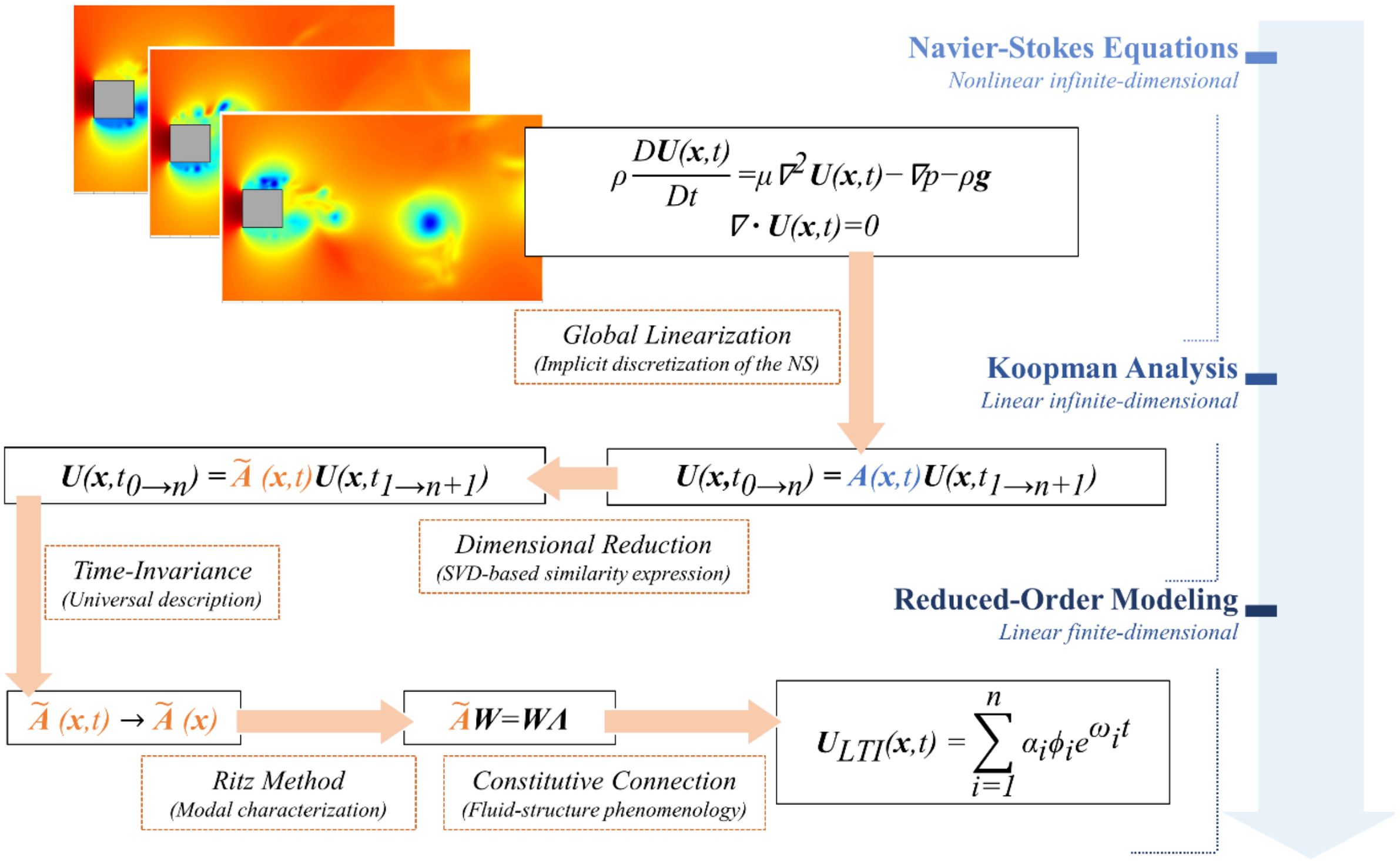

**Fig**.1 A conceptual rundown of the Koopman Linearly Time-Invariant Analysis (Koopman-LTI)

A prototypical configuration—subcritical free-shear flow past a stiff square prism— is selected as the subject of study. This configuration is geometrically simplistic but phenomenologically sophisticated, involving stagnation, forced separation, reattachment, *etc.* [18]–[21]. The four prism walls are behaviorally distinct [22]–[26], while an infinite spanwise length prevents three-dimensional complications [24], [26]–[28]. An infinite stiffness is also prescribed to simplify the bi-directional feedback loop of FSI into a mono-directional, fluid-to-structure scenario, as often adopted in large-scale structures or civil infrastructure [29]–[31]. In essence, this work aims to present the most fundamental yet sufficiently sophisticated case to encourage an intellectual resonance with the readership. The subcritical regime is selected because success on inhomogeneous and anisotropic turbulence would tell a great deal about the technique's generality for a class of fluid systems, if not many stochastic processes elsewhere. To this end, inlet perturbation is spared to avoid synthetic dynamics. $Re=22,000$ is chosen because this shear layer transition regime appeals to phenomenological similitude for a wide range of $Re$ [32], while permitting high-fidelity simulation with validation [19], [33], [34]. This work employs numerical data because whilst error is possible for numerical data, noise contamination is unlikely. Moreover, the fine spatial discretization needed to resolve turbulence guarantees high-dimensional data, whereas field and experimental data are acutely limited by apparatus.

The Large-Eddy Simulation with Near-Wall Resolution (LES-NWR) was employed to simulate the turbulent flow [35], by the Smagorinsky model [36], the Lilly formulation [37], and the dynamic-stress model [38] [39]. **Fig**. 2 illustrates the computational domain and boundary conditions of the simulation. Readers are referred to our previous work [17] for comprehensive numerical details and case validation.

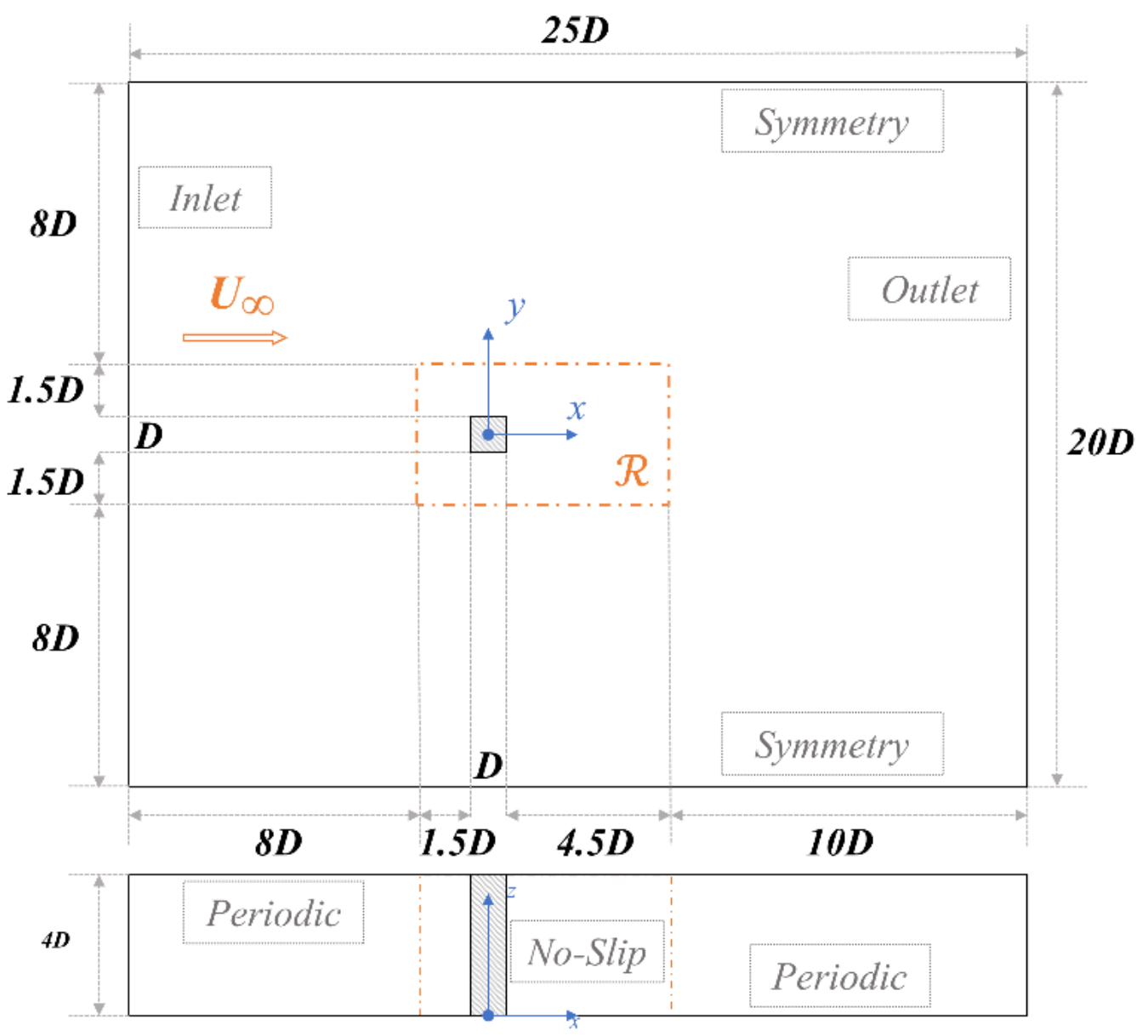

**Fig.** 2 Computational domain and boundary conditions of the LES-NWR-simulated turbulent free-shear flow.

Time invariance of the Koopman operator $\widetilde{A}\ (\boldsymbol{x},t) \rightarrow \widetilde{A}\ (\boldsymbol{x})$ is a critical aspect of the Koopman-LTI, which is pragmatically achieved in two steps. The first is the sampling of mean-subtracted data in the full statistical stationary state, which is corroborated by global and local statistics of the flow. The second is the independence of data sampling, which is guaranteed by reaching the *Stabilization* state when using the DMD to approximate the Koopman modes. The procedure and criteria for reaching time invariance are presented in [17], [40]. In this work, the five input data sequences are the pressures of the four prism walls and the flow field. They are sampled and analyzed independently without any algorithmic communication and information about physics.

In terms of results, we first examine the stability of the Koopman-LTI systems. **Fig**. 3 displays the $\Re$-$\Im$ spectrum, in which each Ritz pair corresponds to DMD conjugates or poles on the Z-space. The poles lie close to the $\Re^2 + \Im^2 = 1$ circle, exhibiting near-perfect oscillatory behaviors and stellar stability, while associated with unique frequencies in continuous time. The stability shows the Koopman-LTI excels in modeling periodic behaviors because, reminded of the Richardson-Kolmogorov's notion, turbulence is made up of eddies of different sizes, which translate precisely into motions of various periodicities in the wavenumber space [35]. Quantitatively, the maximum $l_2$ normalized reconstruction error (Eq. 4.3.1.1 in [17]) of all five data sequences is capped at $10^{-11}$, which is essentially a numerical zero.

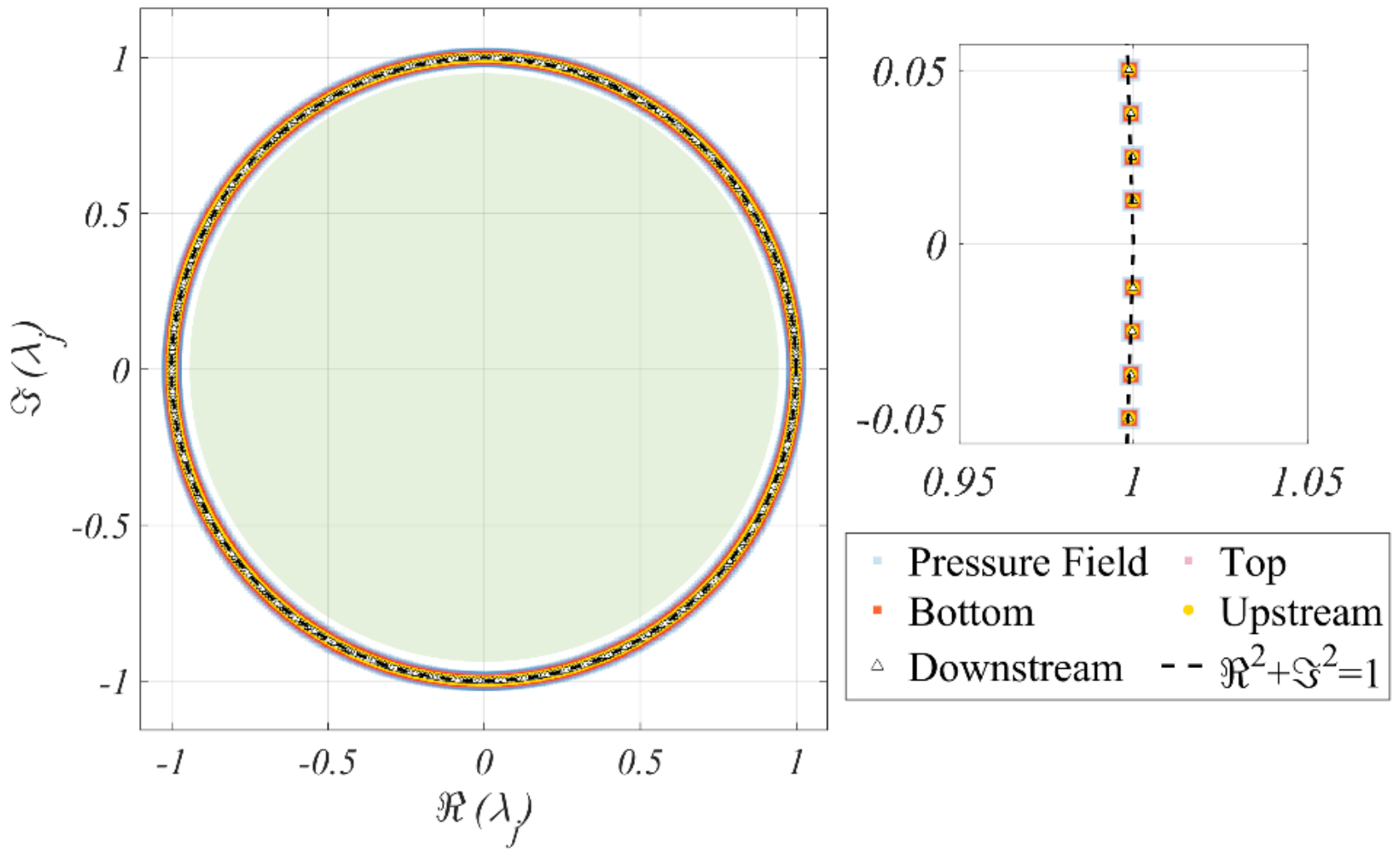

**Fig** 3. The $\Re$-$\Im$ spectra (markers) and *Region of Convergence* (mint green) of the Koopman-LTI systems

**Fig**. 3 also illustrates the *Region of Convergence* (ROC) in mint green. The ROC contains the origin and is enclosed by poles and the $\Re^2 + \Im^2 = 1$ circle, characterizing an anti-causal system that depends only on future inputs. This is concrete evidence of time-invariance. Furthermore, the poles of the five systems are identical, as they superpose perfectly onto one another. **This observation is of tremendous importance: despite the sampling and analytical autonomy, a configuration-wise universal Koopman-LTI system exists, implying the governance of a consistent set of physical principles.**

The only difference between the five Koopman-LTI systems is the ranking of modal dominance by $\alpha_i$ (**Fig.** 4). To this end, the pressure field and pressures on the top (*BC*), down (*AD*), and upstream (*AB*) walls share similar behavioral trends, whereas the downstream (*CD*) wall exhibits smaller differences between the leading mode and the others. This echoes with the findings of Unal and Rockwell on base pressure [41] that results from tumultuous vortical and entrainment activities (both stream- and span-wise) in the undulating near-wake, whereas those of the other three walls are generally dictated by separation and the shear layers. The distribution of the modal amplitude also resembles the discrete Fourier transform, which has been confirmed by [13], [14].

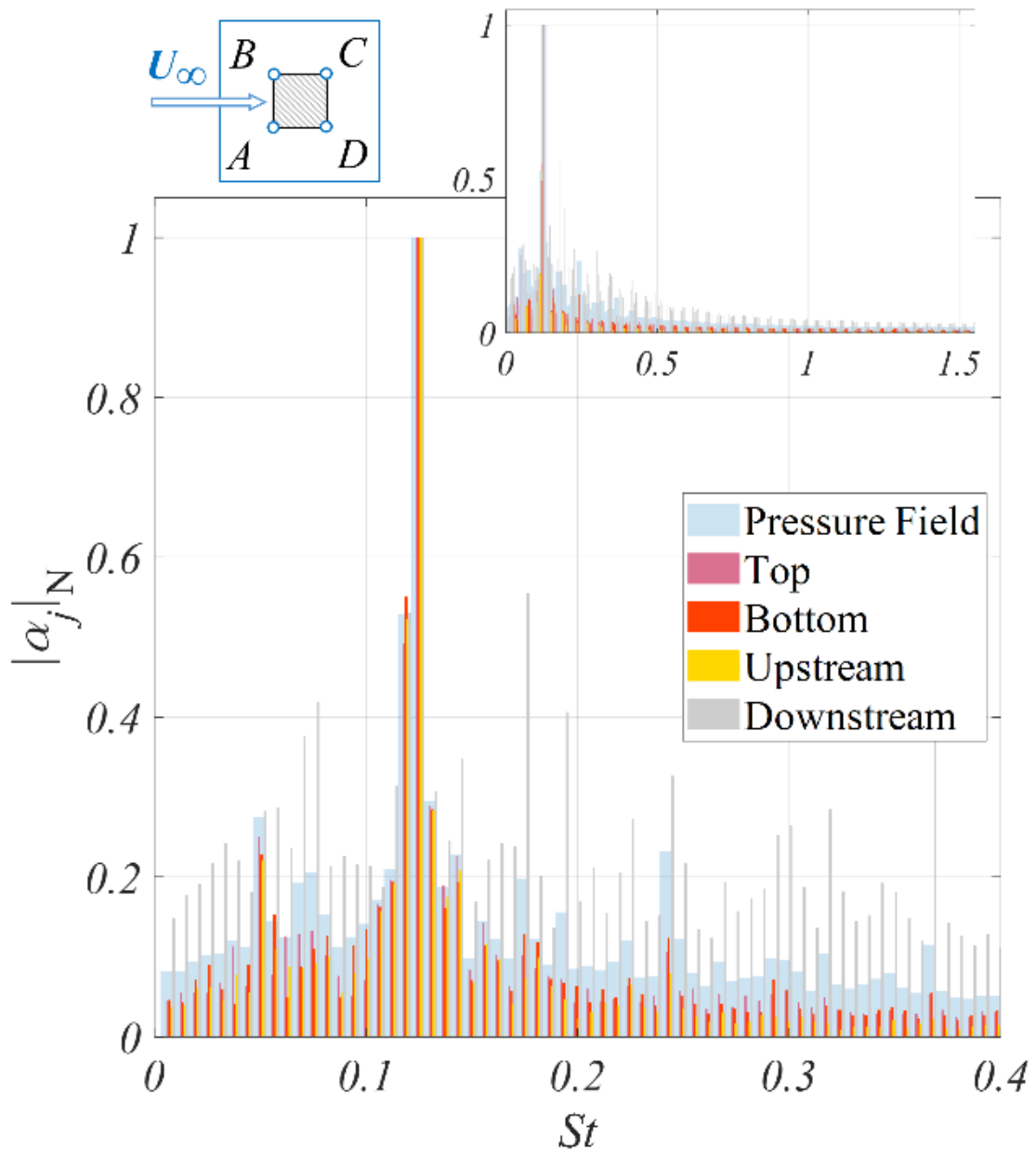

**Fig.** 4 Normalized leading coefficients $\alpha_j$ versus the Strouhal number, showing spatiotemporal dominance and distribution of modal frequency

**Fig.** 4 also illustrates how the **Koopman-LTI decomposes nonlinearly entwined raw data into distinct constituents**. By mathematical formulation, an input sequence containing $N$ snapshots produces $N$ temporally orthogonal but not spatially mixed Ritz pairs. It is analogous to performing a spectral analysis for the Proper Orthogonal Decomposition (POD) modes, after which the energy content of a POD mode distributes across several DMD modes. Moreover, since the turbulence spectrum is continuous, increasing $N$ essentially refines the spectral discretization. As $N \rightarrow \infty$ and $t \rightarrow \infty$, in principle $\widetilde{\boldsymbol{A}}\,(\boldsymbol{x},t) \rightarrow$ the Navier-Stokes Equations, and the Z-transform converges to the Laplace-transform $\mathcal{Z}(\boldsymbol{U}[n]) \rightarrow \mathcal{L}(\boldsymbol{U}(t))$.

The ensuing analysis is performed for the 30 most dominant Koopman-LTI modes. However, to preserve literary concision, only Modes 1 and 4 are presented in **Figs.** 5 and 6, respectively. Matching modal frequencies permit straightforward connections between the phenomenology of the fluid and structure for every time instant ($t^*=\Delta t U_\infty/D$ reduced time). Clearly, Mode 1 describes an alternatingly antisymmetric phenomenon that stems from flow separation, which triggers periodic bulging motions of the separation bubble before synchronously shedding into the near-wake. This field phenomenon is responsible for the periodic shift of positive and negative pressures on the cross-wind walls, and the axisymmetric alteration on the downstream wall. Given $St_1 = 0.124$, the mechanism described by Mode 1 is certainly a major component of the Bérnard-Kármán vortex shedding [42].

Mode 4 ($St_4 = 0.0497$) describes a very different scenario. This low-frequency activity appeals to the stretching of vortices. The pattern is unsynchronized and asymmetric, in which a dominant structure extends from the prism base along the central axis, and another weaker one forms towards either side. Structure-wise, while the cross-wind walls are subjected to alternating pressures and clear reattachment, the prism base experiences an intense pressure gradient radiating from a central location. This information can be of great value. In a hypothetical scenario, say for a building, to alleviate pressure extremities on the downstream façade, one shall certainly try to exterminate the phenomenon described by Mode 4.

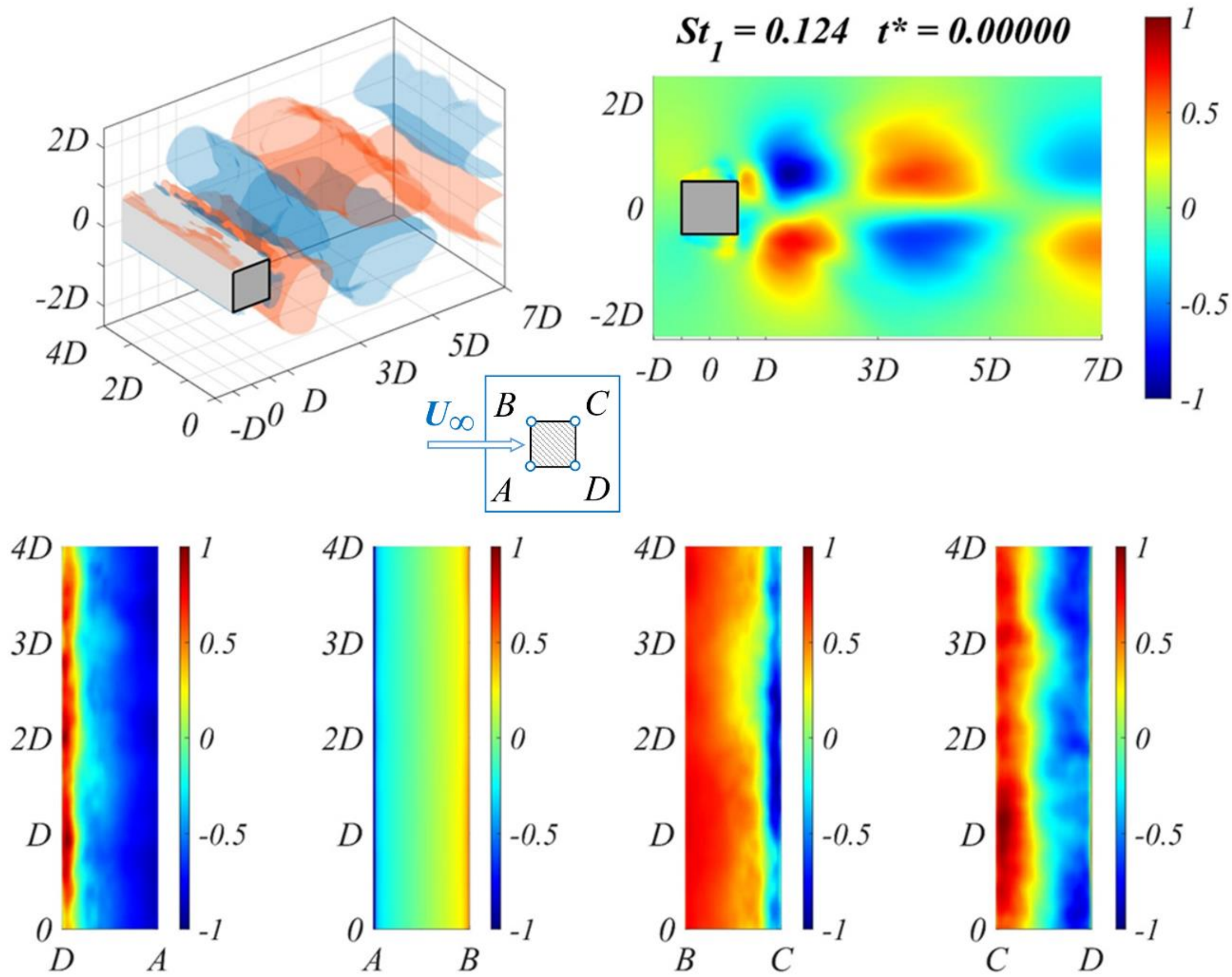

**Fig**. 5 Normalized mode shape (-1 to 1) of Mode 1 ($St_1$ = *0.124*) at *t*=0 (generic)*. Top left: iso-surfaces ±*0.25* of the pressure field. Top right: slice at mid-prism-span. Bottom from left to right: frequency matching mode shapes of the bottom, upstream, top, and downstream walls, respectively. (Multimedia file slowed by a factor of 500)

At this point, the implications of this work are lucid. **The Koopman-LTI analysis permits direct connections between the phenomenology of fluid and structure.** While the link can be somewhat established by other techniques, like spectral density or wavelet analysis, the visualization is generally unavailable. Other techniques like the POD grants modal isolation and visualization, but without the frequency similitude, any speculation that associates the field and walls is unwarranted. Therefore, the Koopman-LTI offers brand-new, physics-revealing insights into fluid-structure interaction. In terms of engineering application, one can now pinpoint the exact excitation mechanism(s) to target when trying to amplify or alleviate a specific structural behavior, or vise versa.

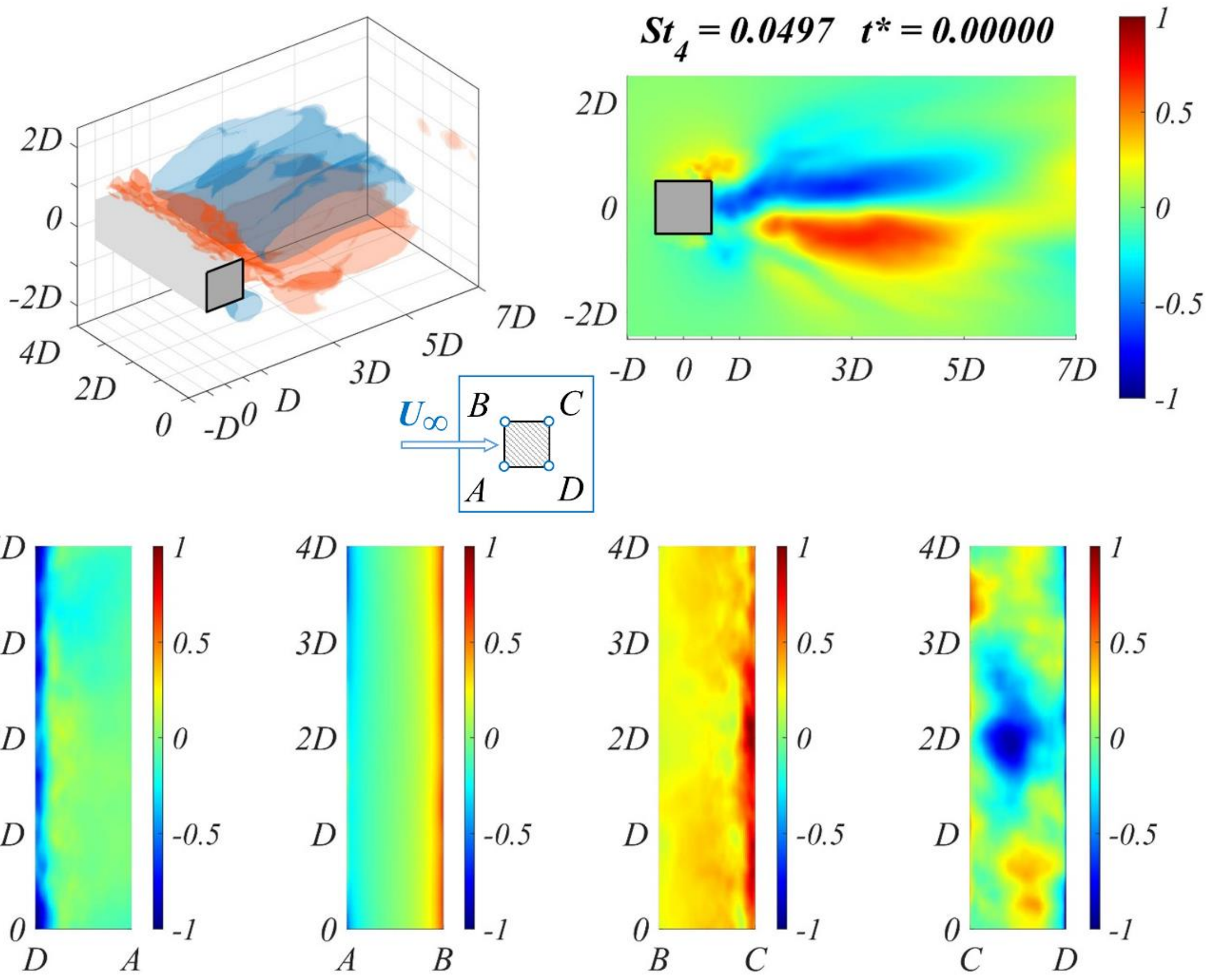

**Fig.** 6 Normalized mode shape (-1 to 1) of Mode 4 ($St_4$ = *0.0497*) at $t^*=0$ *(generic)*. Top left: iso-surfaces ±*0.25* of the pressure field. Top right: contour sliced at mid-prism-span. Bottom from left to right: frequency matching mode shapes of the bottom, upstream, top, and downstream walls, respectively. (Multimedia file slowed by a factor of 500)

Some limitations of the Koopman-LTI also need to be addressed. First, since each mode represents a group of similar-size eddies moving at similar convective velocities, the number of modes can essentially extend to infinity. Consequently, data management becomes a strenuous task for ordinary computational platforms. Second, the spatiotemporal index for mode dominance ranking must strike a proper balance in the weighting of energy and time. As the authors [40] and other researchers [12], [17], [43] have shown, certain low-energy states may have high dynamical relevance. To date, although several criteria have been proposed [12], [44]–[47], there is yet a universal consensus. Third, errors will always ensue from the Koopman linearization. Although minuscule herein, they may be significant for more sophisticated systems. Finally, the discussions herein only apply to stationary flows. Applicability to transient flows demands future investigation.

We give a special thanks to the IT Office of the Department of Civil and Environmental Engineering at the Hong Kong University of Science and Technology. Its support for the installation, testing, and maintenance of our high-performance servers are indispensable for the current project.

The authors declare that they have no conflict of interest.

The datasets generated during and/or analyzed during the current work are restricted by provisions of the funding source but are available from the corresponding author on reasonable request.

The custom code used during and/or analyzed during the current work is restricted by provisions of the funding source.

All authors contributed to the study's conception and design. Funding, project management, and supervision were performed by Tim K.T. Tse and Zengshun Chen. Material preparation, data collection, and formal analysis were led by Cruz Y. Li and Zengshun Chen, and assisted by Asiri Umenga Weerasuriya, Xuelin Zhang, Yunfei Fu, and Xisheng Lin. The first draft of the manuscript was written by Cruz Y. Li and all authors commented on previous versions of the manuscript. All authors read, contributed, and approved the final manuscript.

All procedures performed in this work were in accordance with the ethical standards of the institutional and/or national research committee and with the 1964 Helsinki declaration and its later amendments or comparable ethical standards.

Informed consent was obtained from all individual participants included in the study.

Publication consent was obtained from all individual participants included in the study.